\documentclass[fleqn,twoside]{article}
\usepackage{espcrc2}

\readRCS
$Id: espcrc2.tex,v 1.2 2004/02/24 11:22:11 spepping Exp $
\usepackage{graphicx}
\usepackage[figuresright]{rotating}

\newcommand{\AmS}{{\protect\the\textfont2
  A\kern-.1667em\lower.5ex\hbox{M}\kern-.125emS}}

\title{ Progress in the Small $x$ Resummation\\ of the Singlet Anomalous Dimension}

\author{Guido ALTARELLI, \address[MCSD]{CERN, Department of Physics, Theory Division\\ 
        CH-1211 Geneva 23, Switzerland}%
        Richard D. BALL\address{School of Physics, University of Edinburgh \\
        Edinburgh EH9 3JZ, Scotland}
        and
        Stefano FORTE\address[MCSD]{Dipartimento di Fisica, Universit\'a di Milano and \\
	INFN, Sezione di Milano, Via Celoria 16, I-20133 Milan, Italy}}
       
\begin{document}
\begin{titlepage}
\setcounter{page}{0}
\topmargin 2mm
{\large \begin{flushright}
{\tt hep-ph/0407153}\\
CERN-PH-TH/2004-127\\
Edinburgh 2004/13\\
IFUM-801/FT\\
\end{flushright}
\begin{center}
{\bf\Large Progress in the  Small $x$ Resummation}\\ 
{\bf\Large of the Singlet Anomalous Dimension}
\vskip 24pt
Guido Altarelli\\
\vskip 6pt
{\it CERN, Department of Physics, Theory Division}\\
{\it CH-1211 Gen\`eve 23, Switzerland}\\
\vskip 9pt
Richard D.~Ball\\
\vskip 6pt
{\it School of Physics, University of Edinburgh}\\
{\it  Edinburgh EH9 3JZ, Scotland}\\
\vskip 9pt
Stefano Forte\\
\vskip 6pt
{\it Dipartimento di  Fisica, Universit\`a di
Milano and}\\
{\it INFN, Sezione di Milano, Via Celoria 16, I-20133
Milan, Italy}
\vskip 45pt
\end{center}
\centerline{\bf ABSTRACT}
\bigskip\noindent
{\advance\leftskip by 36truept\advance\rightskip by 36truept
We summarize our
recent results on the small $x$ resummation of the singlet 
anomalous dimension. 
We recall the main features of our approach, and briefly
describe some work in progress on the inclusion of subleading
corrections to it.

\smallskip
}

\begin{center}
\vskip 30pt
Presented by G.A. at Loops\&Legs 2004\\
{ Zinnowitz, April 2004}\\
{\it To be published in the proceedings}\\
\end{center}
\bigskip
\vfill
\begin{flushleft}
CERN-PH-TH/2004-127\\
June 2004\\
\end{flushleft}}
\end{titlepage}

\def\noblackbox{\overfullrule=0pt}
\hyphenation{anom-aly anom-alies coun-ter-term coun-ter-terms}
\def\inv{^{\raise.15ex\hbox{${\scriptscriptstyle -}$}\kern-.05em 1}}
\def\dup{^{\vphantom{1}}}
\def\Dsl{\,\raise.15ex\hbox{/}\mkern-13.5mu D} 
\def\dsl{\raise.15ex\hbox{/}\kern-.57em\partial}
\def\del{\partial}
\def\Psl{\dsl}
\def\tr{{\rm tr}} \def\Tr{{\rm Tr}}
\font\bigit=cmti10 scaled \magstep1
\def\biglie{\hbox{\bigit\$}} 
\def\lspace{\ifx\answ\bigans{}\else\qquad\fi}
\def\lbspace{\ifx\answ\bigans{}\else\hskip-.2in\fi} 
\def\boxeqn#1{\vcenter{\vbox{\hrule\hbox{\vrule\kern3pt\vbox{\kern3pt
        \hbox{${\displaystyle #1}$}\kern3pt}\kern3pt\vrule}\hrule}}}
\def\mbox#1#2{\vcenter{\hrule \hbox{\vrule height#2in
                \kern#1in \vrule} \hrule}}  
%
\def\CAG{{\cal A/\cal G}} \def\CO{{\cal O}} 
\def\CA{{\cal A}} \def\CC{{\cal C}} \def\CF{{\cal F}} \def\CG{{\cal G}}
\def\CL{{\cal L}} \def\CH{{\cal H}} \def\CI{{\cal I}} \def\CU{{\cal U}}
\def\CB{{\cal B}} \def\CR{{\cal R}} \def\CD{{\cal D}} \def\CT{{\cal T}}
\def\e#1{{\rm e}^{^{\textstyle#1}}}
\def\grad#1{\,\nabla\!_{{#1}}\,}
\def\gradgrad#1#2{\,\nabla\!_{{#1}}\nabla\!_{{#2}}\,}
\def\ph{\varphi}
\def\psibar{\overline\psi}
\def\om#1#2{\omega^{#1}{}_{#2}}
\def\vev#1{\langle #1 \rangle}
\def\lform{\hbox{$\sqcup$}\llap{\hbox{$\sqcap$}}}
\def\darr#1{\raise1.5ex\hbox{$\leftrightarrow$}\mkern-16.5mu #1}
\def\lie{\hbox{\it\$}} 
\def\chipp{\kappa}
\def\ha{{1\over2}}
\def\half{{\textstyle{1\over2}}} 
\def\roughly#1{\raise.3ex\hbox{$#1$\kern-.75em\lower1ex\hbox{$\sim$}}}
\hyphenation{Mar-ti-nel-li}
\def\Im{\,\hbox{Im}\,}
\def\Re{\,\hbox{Re}\,}
\def\tr{\,{\hbox{tr}}\,}
\def\Tr{\,{\hbox{Tr}}\,}
\def\det{\,{\hbox{det}}\,}
\def\LLx{leading log~$x$}
\def\Det{\,{\hbox{Det}}\,}
\def\ker{\,{\hbox{ker}}\,}
\def\dim{\,{\hbox{dim}}\,}
\def\ind{\,{\hbox{ind}}\,}
\def\1{\;1\!\!\!\! 1\;}
\def\sgn{\,{\hbox{sgn}}\,}
\def\mod{\,{\hbox{mod}}\,}
\def\eg{{\it e.g.}}
\def\ie{{\it i.e.}}
\def\viz{{\it viz.}}
\def\etal{{\it et al.}}
\def\rhs{right hand side}
\def\lhs{left hand side}
\def\toinf#1{\mathrel{\mathop{\sim}\limits_{\scriptscriptstyle
{#1\rightarrow\infty }}}}
\def\tozero#1{\mathrel{\mathop{\sim}\limits_{\scriptscriptstyle
{#1\rightarrow0 }}}}
\def\frac#1#2{{{#1}\over {#2}}}
\def\half{\hbox{${1\over 2}$}}
\def\threehalf{\hbox{${3\over 2}$}}
\def\third{\hbox{${1\over 3}$}}
\def\quarter{\hbox{${1\over 4}$}}
\def\smallfrac#1#2{\hbox{${{#1}\over {#2}}$}}
\def\neath#1#2{\mathrel{\mathop{#1}\limits{#2}}}
\def\pbp{\bar{\psi }\psi }
\def\vevpbp{\langle 0|\pbp |0\rangle }
\def\tr{{\rm tr}}\def\Tr{{\rm Tr}}
\def\eV{{\rm eV}}\def\keV{{\rm keV}}
\def\MeV{{\rm MeV}}\def\GeV{{\rm GeV}}\def\TeV{{\rm TeV}}
\def\blackbox{\vrule height7pt width5pt depth2pt}
\def\matele#1#2#3{\langle {#1} \vert {#2} \vert {#3} \rangle }
\def\VertL{\Vert_{\Lambda}}\def\VertR{\Vert_{\Lambda_R}}
\def\Real{\Re e}\def\Imag{\Im m}
\def\SZP{\hbox{S0$'$}}\def\DZP{\hbox{D0$'$}}\def\DMP{\hbox{D-$'$}}
\def\MS{\hbox{$\overline{\rm MS}$}}
\def\ms{\hbox{$\overline{\scriptstyle\rm MS}$}}
\def\QMS{Q$_0$\MS}
\def\QDIS{Q$_0$DIS}
\catcode`@=11 
\def\slash#1{\mathord{\mathpalette\c@ncel#1}}
 \def\c@ncel#1#2{\ooalign{$\hfil#1\mkern1mu/\hfil$\crcr$#1#2$}}
\def\lsim{\mathrel{\mathpalette\@versim<}}
\def\gsim{\mathrel{\mathpalette\@versim>}}
 \def\@versim#1#2{\lower0.2ex\vbox{\baselineskip\z@skip\lineskip\z@skip
       \lineskiplimit\z@\ialign{$\m@th#1\hfil##$\crcr#2\crcr\sim\crcr}}}
\catcode`@=12 
\def\twiddles#1{\mathrel{\mathop{\sim}\limits_
                        {\scriptscriptstyle {#1\rightarrow \infty }}}}
\def\PR{{\it Phys.~Rev.~}}
\def\PRL{{\it Phys.~Rev.~Lett.~}}
\def\NP{{\it Nucl.~Phys.~}}
\def\NPBPS{{\it Nucl.~Phys.~B (Proc.~Suppl.)~}}
\def\PL{{\it Phys.~Lett.~}}
\def\PRep{{\it Phys.~Rep.~}}
\def\AP{{\it Ann.~Phys.~}}
\def\CMP{{\it Comm.~Math.~Phys.~}}
\def\JMP{{\it Jour.~Math.~Phys.~}}
\def\NC{{\it Nuov.~Cim.~}}
\def\SJNP{{\it Sov.~Jour.~Nucl.~Phys.~}}
\def\SPJETP{{\it Sov.~Phys.~J.E.T.P.~}}
\def\ZP{{\it Zeit.~Phys.~}}
\def\JP{{\it Jour.~Phys.~}}
\def\JHEP{{\it Jour.~High~Energy~Phys.~}}
\def\vol#1{{\bf #1}}\def\vyp#1#2#3{\vol{#1} (#2) #3}
\def\dfac{double factorization}
\def\mfac{mass factorization}
\def\efac{energy factorization}
\def\rge{renormalization group equation}
\def\rg{renormalization group}
\def\trge{transverse renormalization group equation}
\def\gdl{\gamma_{\scriptstyle\rm DL}}
\def\lrge{longitudinal renormalization group equation}
\def\pdf{parton distribution function}
\def\as{\alpha_s}
\def\bas{a}
\def\asmu{\alpha_s(\mu^2)}
\def\asQ{\alpha_s(Q^2)}
\def\asS{\alpha_s(S^2)}
\def\Lam{\Lambda}
\def\muderiv{\mu^2\frac{\partial}{\partial\mu^2}}
\def\Qderiv{Q^2\frac{\partial}{\partial Q^2}}
\def\Sderiv{S^2\frac{\partial}{\partial S^2}}
\def\Ai{\hbox{Ai}}
\def\ash{\widehat\alpha_s}
\def\rgederiv{\muderiv+\beta(\asmu)\frac{\partial}{\partial\alpha_s}}
\noblackbox




\begin{abstract}
We summarize our
recent results on the small $x$ resummation of the singlet 
anomalous dimension. 
We recall the main features of our approach, and briefly
describe some work in progress on the inclusion of subleading
corrections to it.
\vspace{1pc}
\end{abstract}

\maketitle
The understanding of scaling violations for deep inelastic
structure functions at small $x$ has been characterized for some time by
an 
apparent contradiction
between theoretical expectations and experimental results.
On the one hand, new effects beyond the fixed order perturbative
approximation to
anomalous dimensions are expected to become important at
small $x$.  On the other hand, no significant deviation from a standard
next--to--leading order perturbative treatment of  scaling
violations of
 structure function data  has been found. Thanks to a body of
 theoretical work by our group~\cite{newabf}, and, along similar lines, by 
the authors of ref.~\cite{newciaf} (see also~\cite{thorne}), 
the origin of this situation is now essentially
understood.

Perturbative anomalous dimensions have been recently~\cite{VMM}
computed up to NNLO:
\begin{equation}
\gamma(N,\alpha_s)=\alpha_s \gamma_0(N)~+~\alpha_s^2
\gamma_1(N)~+~\alpha_s^3
\gamma_2(N)~~\dots .
\label{gammadef}
\end{equation}
This perturbative expansion is not reliable at small $x$, when $\alpha
\ln{1\over x}\sim 1$, as is already the case in the HERA region. 
The
problem is how to use the information contained in the BFKL kernel
to resum it in such a way that the improved splitting function
remains a good approximation down to small values
of $x$. 
This can be accomplished~\cite{newabf} by exploiting the
fact that the solutions of the BFKL and GLAP
equations coincide at leading twist if their respective evolution kernels are
related by ``duality''~\cite{dual}.  In the fixed coupling limit
the duality relation is simply given by
\begin{equation}
\chi(\gamma(N,\as),\as)=N,
\label{dualdef}
\end{equation}
where $\chi(\alpha_s,M)$  is the BFKL kernel, which has been computed to
NLO:
\begin{equation}
\chi(M,\alpha_s)=\alpha_s \chi_0(M)~+~\alpha_s^2 \chi_1(M)~+~\dots .
\label{chidef}
\end{equation} 
The splitting function will then contain
 all terms of order
$(\alpha_s \log{1/x})^n$, derived from
$\chi_0(M)$, and of order
$\alpha_s(\alpha_s \log{1/x})^n$, derived from $\chi_1(M)$.

The small $x$ behaviour of the recently computed NNLO anomalous
dimensions
displays two undesirable features which characterize small $x$
resummations. First, higher order corrections to the anomalous
dimension are very large at small $x$: the NNLO/NLO ratio grows at
small $x$. Second, the dominant small
$x$ terms do not   approximate well the anomalous dimension even at
rather  small $x$. In
other words, resummation effects appear to be large, and the resummed
expansion does not converge well.

We now understand how to treat both problems.
The problem of the poor behaviour of the
resummed  expansion of the anomalous dimension is
cured  
if the small-$x$ resummation is
combined with the standard resummation of collinear singularities, by
constructing a `double-leading' perturbative
expansion. 
This double leading expansion resums the collinear
 poles at $M=0$ in the expansion~(3), enabling the
 imposition of the physical requirement of momentum
conservation: $\gamma(1,\alpha_s)=0$, whence
\begin{equation}
\chi(0,\alpha_s)=1.
\label{mom}
\end{equation}

The  problem of the excessive size of resummation corrections is
solved by a full treatment of the running coupling:
once running coupling effects are properly included in
the improved anomalous dimension, the
asymptotic behavior near $x=0$ is much softened with respect to the 
Lipatov exponent which characterizes the fixed-coupling resummation. 
Hence, the corresponding
dramatic rise of structure functions at small $x$, which is 
ruled out phenomenologically, is replaced by a milder rise. 

We now briefly
recall our improved splitting function, referring the reader to the
original papers~\cite{newabf,StPeter} 
for a full discussion.
Assuming that one only knows $\gamma_0(N)$, $\gamma_1(N)$ and
$\chi_0(M)$, 
the improved anomalous dimension has the following expression:
\begin{eqnarray}
&&\gamma_I^{NL}(\as, N) = \Big[\as\gamma_0(N)+ \as^2 \gamma_1(N) \nonumber\\
&&+\gamma_s(\smallfrac{\as}{N}) -\smallfrac{n_c\as}{\pi N}\Big]
+\gamma_A(c_0,\as,N)\nonumber\\&&-\half +
\sqrt{\smallfrac{2}{\kappa_0\as}[N-\as
c_0]} +\quarter\beta_0\as(1+\frac{\as}{N} c_0)\nonumber\\
&&-\rm{mom.~sub.}\nonumber\\
\end{eqnarray}
The first group of terms on
the right-hand side, within square brackets,
is the sum of the leading order of the double leading expansion (DL-LO
approximation), plus the next-to-leading correction $\gamma_1$ to the
standard GLAP anomalous dimension: namely, it is the sum of
the NLO perturbative
term
$\as\gamma_0(N)+\as^2\gamma_1(N)$ plus the power series of terms
$(\as/N)^n$ in $\gamma_s(\smallfrac{\as}{N})$, obtained
from $\chi_0$ using eq.~(2), with subtraction of the order
$\as$ term to avoid double counting. 
In the second
line, the ``Airy'' anomalous dimension
$\gamma_A(c_0,\as,N)$ contains the running coupling resummation, 
and the terms in the third line
subtract the contributions to $\gamma_A(c_0,\as,N)$
which are already included in
$\gamma_s$,
$\gamma_0$ and $\gamma_1$.   
The Airy anomalous dimension $\gamma_A(c_0,\as,N)$ 
is the exact solution of the 
running coupling BFKL equation which corresponds to a quadratic 
approximation to $\chi_0$ near $M=\half$: $\chi_0 \approx
[c_0+\half\kappa_0(M-\half)^2]$. Finally ``$\rm{mom.~sub.}$" is a
subleading subtraction  that ensures exact momentum
conservation $\gamma_I^{NL}(1,\alpha_s)=0$.

The properties of the improved anomalous dimension in this
approximation are the following. In the limit $\as
\rightarrow 0$ with $N$ fixed, $\gamma_I(\as,N)$ reduces to
$\as\gamma_0(N)+\as^2\gamma_1(N)+O(\as^3)$. For  $\as
\rightarrow 0$ with $\as/N$ fixed, $\gamma_I(\as,N)$  reduces to
$\gamma_s(\smallfrac{\as}{N})$, i.e. the leading term of the
small $x$ expansion. Thus the Airy term is subleading in both
limits. In spite of this, its role is very significant because of the
singularity structure of the different terms in
eq.~(5). Indeed,
$\gamma_0(N)$ has a pole at $N=0$, $\gamma_s$ has a branch cut at
$N=\as c_0$, and
$\gamma_A$ has a pole at $N=N_0<\as c_0$, where $N_0$ is the position of the
rightmost zero of the Airy function. The
importance of the Airy term is that the square root term subtracted from
$\gamma_A$ cancels, within the relevant accuracy, the branch cut of $\gamma_s$
at $N=\as c_0$ and replaces the
corresponding asymptotic behaviour at small $x$ with the much softer one from
$\gamma_A$. Note that the quadratic approximation is sufficient to
give the correct asymptotic behaviour up to terms which
are of subleading order in comparison to those included in the double-leading
expression in eq.~(5).

\begin{figure}
\includegraphics[width=.48\textwidth]{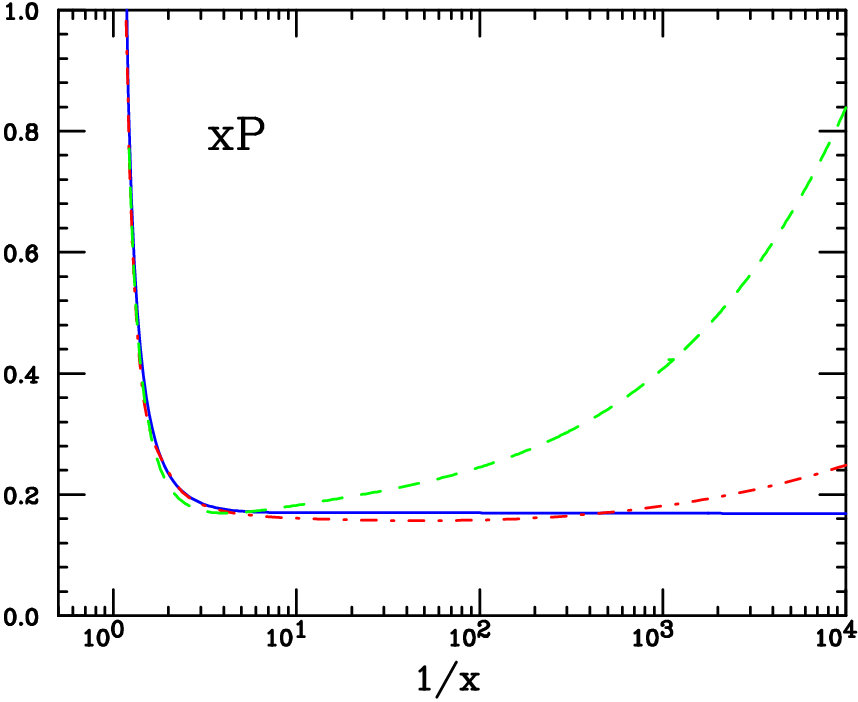}
\caption{The improved splitting function corresponding to 
$\gamma_I^{NL}(\as,N)$ eq.~(5) with
$\as=0.2$ (dot-dashed), compared with those  from the DL-LO
approximation (dashed)   and 
GLAP NLO (solid).}
\end{figure}

The singlet
splitting function obtained from eq.~(5) (for $\as=0.2$) is
shown in Figure~1,
compared with the NLO GLAP kernel and with the DL-LO approximation,
which displays the sharp small-$x$ rise characteristic of the BFKL resummation.
In the region of the HERA
data,  our improved splitting function,  with no free
parameters, closely follows
the  NLO GLAP evolution with a behaviour at
small $x$ which is much softer than that of BFKL. 
It is interesting to
note that the agreement between GLAP and resummed results is
significantly improved by the inclusion in eq.~(5) 
of $\gamma_1$ and the
corresponding double-counting subtraction. This improvement was
already shown in ref.~\cite{newabf} in the anomalous dimension,
and is even more apparent in the splitting function from
ref.~\cite{StPeter} displayed in figure~1.

Given that also $\chi_1$ is in fact known~\cite{FadLip},
it might be worth pursuing a 
full next-to-leading order
resummation. This requires the inclusion in the resummed
anomalous dimension of the terms $\gamma_{ss}$ derived from the NLO
BFKL kernel $\chi_1$, not included in eq.~(5). However, 
the running-coupling resummation is based on a quadratic
approximation of the BFKL kernel about its minimum,
while the
next-to-leading order double-leading anomalous dimension has no
minimum. It is possible to include the next-to-leading corrections
upon the assumption that the minimum is restored by unknown
higher-order corrections, but the results are then affected by
large ambiguities~\cite{newabf}.

The lack of minimum in the next-to-leading order double-leading result
is due to the fact that while the double-leading expansion resums the 
$M=0$ poles of 
$\chi(M)$, the poles at $M=1$ are not eliminated and in particular 
introduce a large negative contribution to  $\chi_1(M)$ 
near $M=1$. The resummation of these $M=1$ poles, however, as
emphasized by the authors of ref.~\cite{newciaf}, can be
accomplished  if we recall that the 
underlying BFKL kernel is symmetric under exchange of the two gluons 
of virtual masses $k_1^2$ and $k_2^2$~\cite{FadLip}. In Mellin space, this
implies that the kernel must be symmetric upon $M\leftrightarrow 1-M$. 
Nevertheless, the DIS kernel remains asymmetric, due
(in the fixed coupling limit)  to
the change of the symmetric variable $s/k_1k_2$ into the
corresponding 
variable $s/Q^2$ appropriate for DIS.  Further asymmetric terms,
 to next-to-leading order, are
due to the running of the coupling.

The kernel $\chi_{\rm DIS}$, dual to the DIS anomalous
dimension, is related to the symmetric one
$\chi_{\sigma}$ through  the implicit equation~\cite{FadLip} 
\begin{equation}
\chi_{\rm DIS}(M+\half\chi_{\rm \sigma}(M))=\chi_{\sigma}(M).
\label{symm}
\end{equation} 
Hence, we can resum the $M=1$ poles by performing the double-leading
resummation of $M=0$ poles of $\chi_{\rm DIS}$, determining the
associated $\chi_\sigma$ through eq.~(6), then symmetrizing it (as
$\chi_\sigma$ must be symmetric), and finally going back to DIS
variables by using eq.~(6) again in reverse.
Using the momentum
conservation  eq.~(4) and eq.~(6), 
it is easy to show that $\chi_\sigma(M)$ is an entire
function of M, with 
$\chi_\sigma(-\half)=\chi_\sigma(\threehalf)=1$. 
                     Since $c=\chi_\sigma(\half)=O(\alpha_s)$, while
                     $\chi_\sigma(M)\sim 2|M|$ as $|M|\to\infty$, $\chi_\sigma$
                     necessarily has a minimum at $M=\half$. Going back
                     to DIS variables, $\chi_{DIS}$ will also have a
                     minimum, albeit slightly distorted.

In practice, at leading order this procedure leads to a kernel
$\chi_{\rm DIS}(M)$ defined as the solution of the implicit equation
\begin{eqnarray}
&&\chi_{\rm DIS}(M)=\chi_s(\frac{\alpha}{M})+
\chi_s(\frac{\alpha}{1-M+\chi_{\rm DIS}(M)})+\nonumber\\
&& +{n_c\alpha \over\pi} \Bigg(\psi(1)+ 
\psi(1+\chi_{\rm DIS}(M))\nonumber\\
&&
-\psi(M) -\psi(1-M+\chi_{\rm DIS}(M)) \nonumber\\
&&
-\frac{1}{M}- 
\frac{1}{1-M+\chi_{\rm DIS}(M)}\Bigg),\nonumber\\
\label{leadimprnl}
\end{eqnarray} 
to be compared to the usual DL result
\begin{equation}
\chi_{\rm
DL}(M)=\chi_s(\frac{\alpha}{M})+{\alpha\over\pi}
\left(\chi_0(M)-{n_c\over M}\right),
\label{DL}
\end{equation}
       where $\chi_s(\alpha/M)$ is the kernel dual to the leading order
                   GLAP anomalous dimension:
                   $M=\alpha\gamma_0(\chi_s(\alpha/M))$.           
The kernel $\chi_{\rm DIS}(M)$ defined by 
eq.~(7) has the following two features: 1)
it only differs from the double-leading kernel
eq.~(8) by terms which are subleading in the double-leading
expansion; 2) the kernel $\chi_\sigma$ computed from it 
using eq.~(6) is symmetric.  
As a consequence, $\chi_{\rm DIS}(M)$ is free of singularities
                   for all positive $M$, rising to unity at the
                   symmetric momentum conservation point $M=2$ (see
Figure~2).

\begin{figure}
\includegraphics[width=.48\textwidth]{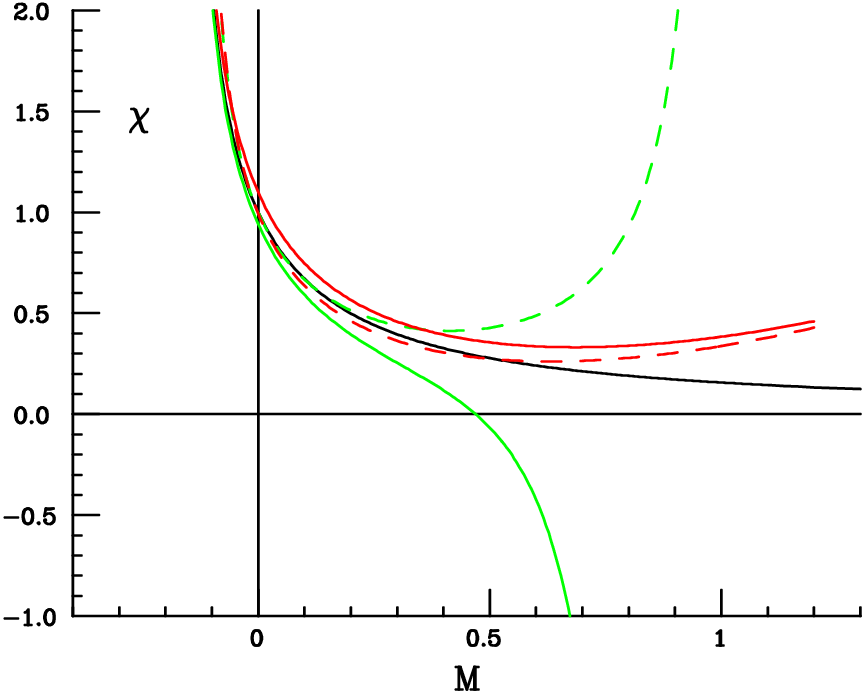}
\caption{The $\chi_{DL}$ functions with and without symmetrisation. From top to
bottom (on the right) the  curves are: 
the LO DL without symmetrisation, the LO+NLO after symmetrisation, the
LO after symmetrisation,
the dual of GLAP LO+NLO and the LO+NLO DL without
symmetrisation}
\end{figure}

The procedure can be extended to next-to-leading order, where various
technical complications arise, due to the need to treat consistently
next-to-leading log $Q^2$ terms, specifically those related to the
running of the coupling. 
 In figure 2 we show the $\chi_{DL}$ kernels with and without
symmetrisation compared to the dual of the GLAP LO+NLO anomalous
dimension. The elimination of the bad behaviour of $\chi_1(M)$ near
$M=1$, which is obtained by symmetrisation, results in a much extended
range of $N$ where the resummed $\chi$ follows GLAP, even before
performing the running coupling resummation.

The presence of a
minimum in the symmetrized $\chi$ makes it suitable for the
running-coupling resummation discussed above. 
In fact, one can easily prove that the parameters which
characterize the running coupling resummation, namely the 
values $c$ and $\kappa$
of the function and the curvature at the minimum, are the same for the
functions $\chi_{\rm DIS}$ and $\chi_\sigma$ related by eq.~(6). One
may thus construct a fully resummed next-to-leading order anomalous dimension
akin to eq.~(5), but based on the symmetrized $\chi$. 
The main ambiguities in the result are related to the
treatment of the running of the coupling. 

\begin{figure}
\includegraphics[width=.48\textwidth]{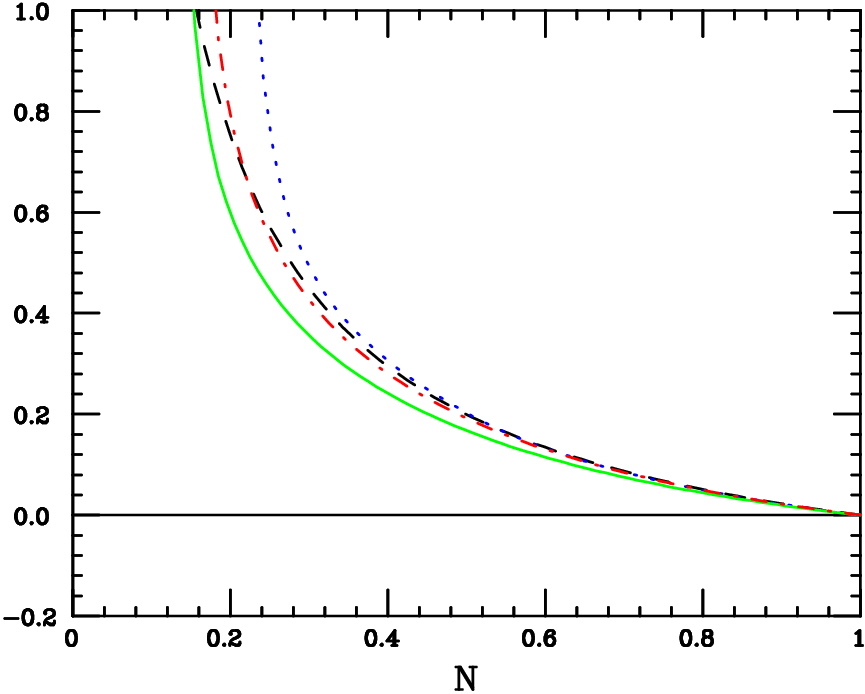}
\caption{Old and new results on the anomalous dimension (see text).}
\end{figure}
The result, for a 'minimal'
running coupling prescription, is
displayed in figure~3. For
clarity, we show an enlargement for the range $0<N<1$: beyond 
$N=1$ the curves
all essentially overlap. The upper (dotted)
curve is the result from eq.~(5), whose associate 
splitting function is displayed in figure 1.
The dashed curve is the LO+NLO GLAP curve. Finally, the lowest (solid)
curve and the central  dot-dashed curve are respectively the LO and
NLO results obtained by applying the running coupling resummation to
the symmetrized kernels. It is apparent that, even though at LO the
new resummed curve somewhat undershoots the GLAP result, the NLO 
resummed curve is even closer to GLAP than the result in
eq.~(5).  A detailed discussion of these results, and
specifically of the ambiguities related to the symmetrization
procedure and running of the coupling, will be presented elsewhere~\cite{prep}.

\end{document}